\documentclass [12PT]{article}

\usepackage{mathtools,amssymb,amsmath}
\usepackage{color,bm,xcolor}
\usepackage{multirow}
\usepackage{bm}
\usepackage{indentfirst}
\usepackage{natbib}
\usepackage{mathrsfs}
\usepackage{enumitem}
\usepackage{soul}
\usepackage{url}
\usepackage[first=0,last=9]{lcg}
\usepackage{color, colortbl}
\usepackage{subcaption}
\usepackage[bottom]{footmisc}
\usepackage[flushleft]{threeparttable}
\definecolor{Gray}{gray}{0.9}

{
	\title{\bf  Lessons Learned from the Bayesian Design and Analysis for the BNT162b2 COVID-19 Vaccine Phase 3 Trial }
	\author{
		 Yuan Ji\thanks{Department of Public Health Sciences, The University of Chicago, Chicago, USA} \ ,
		 Shijie Yuan\thanks{Laiya Consulting Inc., Chicago, USA}
	}
	\date{\today}
}

\newcommand{\textVE}{\text{VE}}

\begin{document}
\maketitle
\begin{abstract}
	The phase III BNT162b2 mRNA COVID-19 vaccine trial is based on a Bayesian design and analysis, and the  main  evidence of vaccine efficacy is presented in Bayesian statistics. Confusion and mistakes are produced in the presentation of the Bayesian results. Some key statistics, such as Bayesian credible intervals, are mislabeled and stated as confidence intervals. Posterior probabilities of the vaccine efficacy are not reported as the main results. We illustrate the main differences in the reporting of Bayesian analysis results for a clinical trial and provide  four  recommendations. We argue that statistical evidence from a Bayesian trial, when presented properly,  is easier to interpret and directly addresses the main   clinical  questions, thereby better supporting regulatory decision making.  We also recommend using abbreviation ``BI'' to represent Bayesian credible intervals as a differentiation to ``CI'' which stands for confidence interval. 
\end{abstract}

\section{Introduction}
The phase III BNT162b2 COVID-19 vaccine trial uses a Bayesian design and analysis method  for the primary efficacy endpoints.  Participants are randomized with a 1:1 ratio to receive the vaccine or the placebo. The primary efficacy endpoints  are based on  $\textVE = 100*(1-IRR)$, in which $IRR$ is computed as the ratio of first confirmed COVID-19 illness rate in the vaccine group to the corresponding illness rate in the placebo group \citep{orenstein1985field}. 

The BNT162b2 vaccine has received emergency use authorization (EUA)   by the US FDA, among other countries and regions. The decision is based on the totality of the evidence, including the efficacy and safety of the vaccine,   from a phase I/II/III trial which   is reviewed by the US FDA and the Vaccine and Related Biologic Product Advisory Committee (VRBPAC). On December 10, 2020, the VRBPAC held a public meeting and voted overwhelmingly to support the EUA.

The safety of the vaccine has been adequately reviewed, which is not the main focus here. Instead, we discuss the evidence of the trial related to the vaccine efficacy. In particular, we show that the presentation of the trial data and  primary efficacy   results are not compatible with the Bayesian statistics, and mistakes are made in the interpretation of the results. While these missteps do not change the   regulatory  decision for the Pfizer/BioNTech BNT162b2 vaccine -- thanks to its superior efficacy -- it is nonetheless critical to discuss and correct the mistakes for future trials that use Bayesian designs and methods. For example, the distinction in the definition and interpretation of the Bayesian credible interval and the frequentist confidence interval must be clearly explained to properly assess the clinical evidence for decision making.

\section{Results}

\cite{polack2020safety} reported the observed trial data and   statistical  analysis results for the BNT162b2 COVID-19 vaccine. The use of  the  Bayesian design and analysis   for the primary efficacy endpoint  appears to   impose some   challenges in the way the statistical results are presented. We identify two main missed opportunities. First, credible intervals are not accurately interpreted and occasionally misrepresented as confidence intervals. Second, the presentation and elaboration of the posterior probabilities of   the  true vaccine efficacy are not emphasized. This is an important point since unlike p-values, posterior probabilities directly answer the clinical question of the trial. A minor point is that the Bayesian models   used in  the trial are not clearly described. We reconstruct the model and reproduce the main efficacy results in \cite{polack2020safety}. We also propose an alternative Bayesian model that may be more compatible of the statistical sampling scheme of the design.   Statistical program for these analyses are presented in the Supplemental Material.  

\subsection{Credible Interval}

The 95\% credible intervals in Table 2 of \cite{polack2020safety} are mistakenly labeled as confidence intervals in the text (the EFFICACY paragraph on page 8 of the paper). 
For example, a 95\% credible interval of (90.3, 97.6) in Table 2 is called a “95\% confidence interval”. 

The main difference between confidence and credible intervals 
 lies in  their interpretations  and  the  evidence the two  intervals   
 represent. In particular, a 95\% credible interval $(a,b)$ for VE means   that  given the data, the probability that VE is greater than $a$ but smaller than $b$ is 0.95. In contrast, a 95\% confidence interval $(c, d)$ for VE is interpreted as the following: were the vaccine trial repeated numerous times, the fraction of the calculated confidence intervals (which would be different for each trial) that encompass the true VE would tend towards 95\% \citep{cox1979theoretical}. Apparently, the Bayesian credible interval directly evaluates the probability of the true vaccine efficacy based on the observed trial data, and the frequentist confidence interval only provides an indirect assessment assuming the vaccine trial were to repeat numerous times.

\subsection{Posterior Probability}

Interestingly, the posterior probabilities of the BNT162b2 VE are only reported as text in the DISCUSSION section of \cite{polack2020safety} but not the RESULTS section. The EFFICACY subsection in RESULTS focused on the reporting of the credible intervals, although the credible intervals were mistakenly written as confidence intervals. 

An advantage of the Bayesian modeling for the BNT162b2 vaccine trial  data  is the ability to report the vaccine efficacy with probabilistic statement, a feature that is not available through p-values. Clinically,   a  direct answer to the vaccine efficacy based on clinical trial data is a statement like the following:

`` Given  the observed efficacy data, there is a probability of $Y$ that the true vaccine efficacy is greater than $X$.''

\noindent For the BNT162b2 trial, when $X=30\%$, $Y$ is greater than 0.9999, and when $X=90\%$, $Y=0.98.$''


\section{Discussion}

\subsection{Recommended Statistical Reporting for a Bayesian Trial}

Bayesian results provide direct answers to clinical questions. To see this, we list   four  recommended Bayesian reporting elements for clinicians and decision makers.

\begin{figure}[!htbp]
	\begin{subfigure}{0.5\textwidth}
		\includegraphics[width=\textwidth]{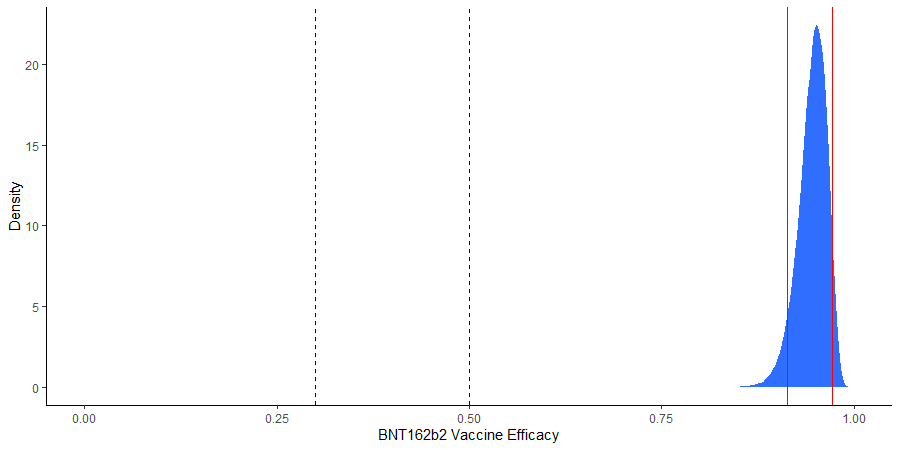}
		\subcaption{{\scriptsize BNT162b2 VE in participants without evidence of infection}}
	\end{subfigure}
	\begin{subfigure}{0.5\textwidth}
		\includegraphics[width=\textwidth]{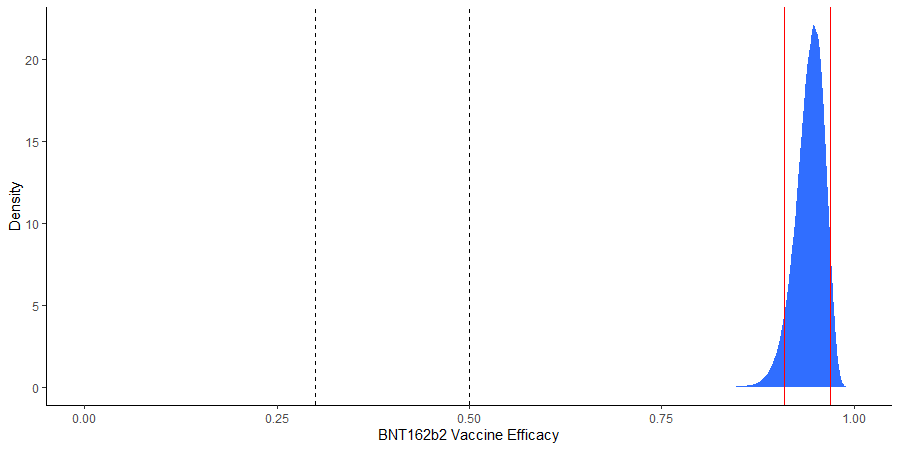}
		\subcaption{{\scriptsize BNT162b2 VE in participants with and without evidence of infection}}
	\end{subfigure}
	\caption{The posterior distribution of the  BNT162b2 VE based on the beta-binomial model in \cite{polack2020safety}. The blue curve is the posterior density of VE. The red lines are the 95\% credible intervals. The two dotted lines  represent  the two VE thresholds, 0.3 and 0.5 mentioned in the FDA guidance for COVID-19 efficacy. Specifically, a vaccine must exhibit observed VE of 0.5 and the lower bound of the 95\% confidence interval must be greater than 0.3, in order to be considered for authorization.}
	\label{fig:posdis-VE}
\end{figure}

\begin{enumerate}[resume,label={\arabic*)}]
	\item \ul{Report posterior probability of clinical benefits or treatment effects.}  For example, $Pr(\textVE > X \mid data)= Y$ provides a direct assessment of the true vaccine efficacy greater than $X$ with confidence (probability) of $Y$, given the observed trial data. In the case of BNT162b2 trial, $Pr(\textVE > 30\% \mid data) > 0.9999$. This means that with a probability larger than 0.9999 the vaccine efficacy is greater than 30\%. In fact, it can be shown (Supplemental Material) that $Pr(\textVE > 90\% \mid data)= 0.98$, which means with a probability 0.98 that the vaccine efficacy is greater than 90\%. This statement is perhaps much more informative for decision making and reflects the superior efficacy of the vaccine. For example, the statement implies that there is only 2\% chance that the vaccine is less than 90\% efficacious.
	
	\item \ul{Report the Bayesian credible interval (BI) and interpret BI  
          using a probability statement.} For example, in the case of BNT162b2 trial, the 95\% credible interval (90.3, 97.6) of VE means that with 95\% probability, the true vaccine efficacy is greater than 90.3\% and less than 97.6\%, given the observed trial data.  We recommend using abbreviation ``BI'' to represent Bayesian credible interval to distinguish ``CI'' which stands for confidence interval.  
	
	\item \ul{Report posterior distribution  (probability) of treatment effects and overlay it with the regulatory thresholds, if possible.} For example, for the BNT162b2 vaccine trial, Figure \ref{fig:posdis-VE} shows the histogram of VE based on its posterior distribution. It is clear that the vaccine efficacy is much higher than the regulatory thresholds of 0.3 and 0.5 \citep{FDACovid}, with most probability mass pointing to values greater than 0.8.  In the BNT162b2 trial, the posterior probability (pp) that VE is greater than 0.3 or 0.5 is greater than 0.99. 

            \item \ul{Report the complete Bayesian models including the prior distributions and the likelihood functions.} This allows transparency so that the assumptions of the models can be assessed and critiqued.  In Supplemental Material, we present two such models, one reproducing the results in \cite{polack2020safety} and the other with better interpretation.  
\end{enumerate}


\subsection{The Bayesian models and Inference for the BNT162b2 trial}

The details of the Bayesian models and inference used in the BNT162b2 vaccine trial were not reported, either in the protocol or \cite{polack2020safety}. We reproduced the reported Bayesian results in the primary efficacy analysis in \cite{polack2020safety}. See Supplemental Material for detail of our model that reproduced the  results.  It is not the first choice of a Bayesian model that we would use,  however,  since the model  is not compatible with the sampling scheme based on the trial design.  An alternative model that is more natural and compatible to the trial design is presented in Supplemental Material , where the BI from the alternative model is (90.8\%, 97.9\%), which is a bit shorter than the reported BI.   

\bibliographystyle{apalike}
\bibliography{vaccine}

\begin{thebibliography}{}

\bibitem[BioNTech, 2020]{biontech2020study}
BioNTech, S. (2020).
\newblock {Study to describe the safety, tolerability, immunogenicity, and
  efficacy of RNA vaccine candidates against COVID-19 in healthy individuals}.
\newblock {\em ClinicalTrials. gov: NCT04368728}.

\bibitem[Cox and Hinkley, 1979]{cox1979theoretical}
Cox, D.~R. and Hinkley, D.~V. (1979).
\newblock {\em Theoretical statistics}.
\newblock CRC Press.

\bibitem[Orenstein et~al., 1985]{orenstein1985field}
Orenstein, W.~A., Bernier, R.~H., Dondero, T.~J., Hinman, A.~R., Marks, J.~S.,
  Bart, K.~J., and Sirotkin, B. (1985).
\newblock Field evaluation of vaccine efficacy.
\newblock {\em Bulletin of the World Health Organization}, 63(6):1055.

\bibitem[Polack et~al., 2020]{polack2020safety}
Polack, F.~P., Thomas, S.~J., Kitchin, N., Absalon, J., Gurtman, A., Lockhart,
  S., Perez, J.~L., P{\'e}rez~Marc, G., Moreira, E.~D., Zerbini, C., et~al.
  (2020).
\newblock {Safety and efficacy of the BNT162b2 mRNA Covid-19 vaccine}.
\newblock {\em New England Journal of Medicine}.

\bibitem[{US FDA}, 2020]{FDACovid}
{US FDA} (2020).
\newblock {\em Emergency Use Authorization for Vaccines to Prevent COVID-19 --
  Guidance for Industry}.

\end{thebibliography}

\newpage
\section*{Supplemental Material}

\renewcommand{\thefigure}{A.\arabic{figure}}
\renewcommand{\thetable}{A.\arabic{table}}
\setcounter{figure}{0}
\setcounter{table}{0}

\subsection*{Reproducible model}
We  report   the following simple beta-binomial model that reproduces the credible intervals in \cite{polack2020safety}. This model assumes that the number of COVID cases in the vaccine group is sampled as a binomial random variable from the total number of COVID cases in both groups, vaccine and placebo. In mathematics, this means 
$$X \mid N,\theta \sim \text{Bin}(X\mid N, \theta),$$
where $X$ denotes the number of cases in the BNT162b2 group and $N$  the  total number of cases in both groups. Therefore, $(N-X)$ is the number of cases in the placebo group. Here, $\theta$ is interpreted as the probability that an observed COVID case is from the vaccine group and $(1-\theta)$ is the probability that it is from the placebo group,  when a COVID case is observed.   Note that the probability sampling space is restricted to only the COVID cases, not including any non-cases. 

A beta prior Beta(0.700102,1) is proposed for $\theta$ in the BNT162b2 protocol \citep{biontech2020study}. Also, the  trial  protocol assumes $\theta=(1-\textVE)/(2-\textVE)$. This assumption means that $\textVE = 1-\theta/(1-\theta),$ which resembles the definition of $\textVE = 1-IRR.$ However, it is important to note that $\theta$ is not the probability of COVID rate in the vaccine group. 

Following this model, with fixed $N$ and $X$, the posterior distribution of $\theta$ is also a beta distribution,
\begin{equation}\label{eq:pos-theta}
\theta \mid N,X \sim \text{Beta}(0.700102+X,1+N-X).
\end{equation}

For the BNT162b2 trial, $X=8$ and $N=162$ for $\textVE_1$ and $X=9$ and $N=169$ for $\textVE_2$,  where $\textVE_1$ and $\textVE_2$ are the two primary efficacy endpoints of the trial.  Therefore, 
 the credible intervals of VE can be calculated via sampling $\theta$ from its beta posterior distribution (\ref{eq:pos-theta}) 
and are shown in Table \ref{tab:model1res} below. They are identical to the reported credible intervals in \cite{polack2020safety}. 
In addition, using the posterior distribution of $\theta$, we easily calculate the posterior probabilities of VEs. For example, $Pr(\textVE_1 > 30\% | data) > 0.9999$ and $Pr(\textVE_2 > 90\% | data) = 0.98.$ See attached computer program for detail. 

\begin{threeparttable}
	\centering
	\caption{Vaccine Efficacy with the prior of $\theta$.} \label{tab:model1res}
	\begin{tabular}{p{16em}ccc}
		\hline
		\textbf{Efficacy End Point} & \textbf{BNT162b2} & \textbf{Placebo} & \multicolumn{1}{p{10em}}{\textbf{Vaccine Efficacy, \% \newline{} (95\% Credible \newline{} Interval)}} \\
		& No. of Cases \newline{} ($X$) & No. of Cases \newline{} ($N-X$) &  \\ \hline
		\rowcolor{Gray}
		$\textVE_1:$ Covid-19 occurrence at least\newline{}
		\hspace*{2em}7 days after the second dose\newline{}
		\hspace*{2em}in participants without evidence\newline{}
		\hspace*{2em}of infection & 8     & 162   &  95.0\tnote{*}    (90.3, 97.6) \\ \hline
		$\textVE_2:$ Covid-19 occurrence at least\newline{}
		\hspace*{2em}7 days after the second dose\newline{}
		\hspace*{2em}in participants with and those\newline{}
		\hspace*{2em}without evidence of infection & 9     & 169   &  94.6\tnote{*} (89.9, 97.3) \\
		\hline
	\end{tabular}
	\begin{tablenotes}
		\item[*]  The reported values of VE are the observed $\textVE = 1- IRR$, the same as in \cite{polack2020safety}, where $IRR$ is based on the observed COVID cases and sample sizes for both groups. We would recommend reporting the posterior means as well, which are 94.6 and 94.3 for the vaccine and placebo groups, respectively, using the reproducible model. 
	\end{tablenotes}
\end{threeparttable}

\subsection*{An alternative model}
We also propose an alternative model that is more natural and compatible to the trial design. Recall that the design of the BNT162b2 trial first enrolls participants without COVID diagnosis, and then follows them for certain time period to observe disease occurrence. This means that for each of the two groups, vaccine and placebo, a binomial sampling is carried out, assuming a homogeneous disease rate within each group and 
 not considering different surveillance time of each patient.  In particular,  let $p_1$ and $p_2$ denote the probabilities of COVID for the vaccine and placebo groups, respectively. Among the $N_1$ participants in the vaccine group and $N_2$ in the placebo group, let $X_1$ and $X_2$ represent the corresponding numbers of COVID cases.

Since participants in the vaccine and placebo groups are treated and followed independently, we assume the following independent binomial sampling distributions, i.e., 
$$X_i \mid N_i,p_i \sim \text{Bin}(X_i \mid N_i,p_i), \quad i=1,2.$$
Note that by definition, $\textVE = 1-p_1/p_2$. Therefore, a Bayesian model and inference is completed by a prior and posterior distribution of $(p_1, p_2).$

\begin{figure}[!htbp]
	\begin{subfigure}{0.5\textwidth}
		\includegraphics[width=\textwidth]{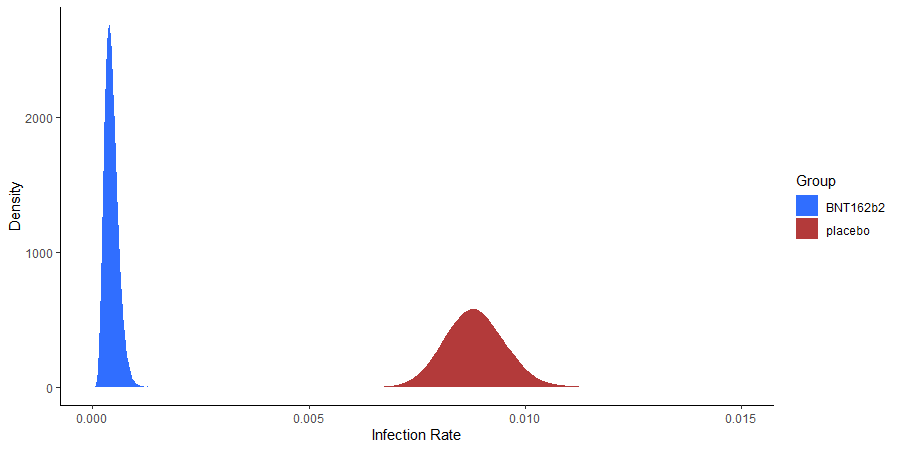}
		\subcaption{{\scriptsize Infection rates of participants without evidence of infection}}
	\end{subfigure}
	\begin{subfigure}{0.5\textwidth}
		\includegraphics[width=\textwidth]{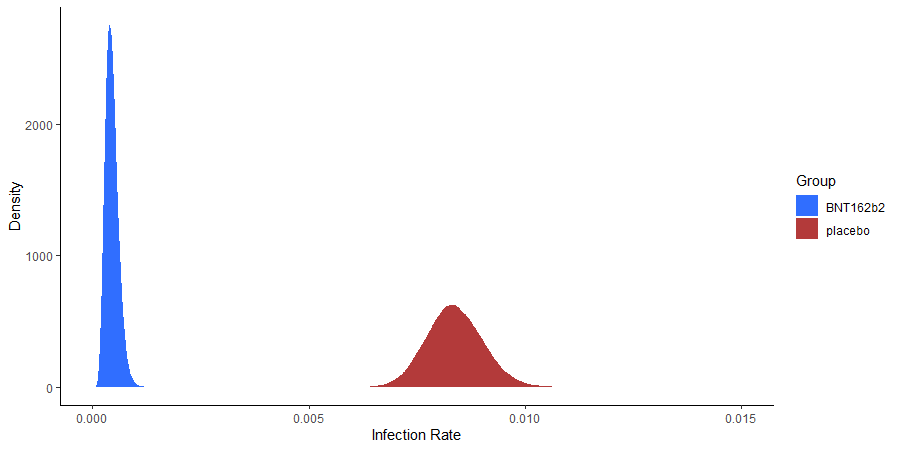}
		\subcaption{{\scriptsize Infection rates of participants with and without evidence of infection}}
	\end{subfigure}
	\caption{The posterior  distributions  of the infection rates for the  BNT162b2  vaccine and placebo groups based on the alternative model.  The blue curves indicate that the vaccine is highly efficacious relative to the placebo, indicated by the red curves. }
	\label{fig:ir-dis}
\end{figure}

We assume that $p_1$ and $p_2$ follow improper and independent prior distributions. In other words, $f(p_i) \sim  1$. This prior leads to proper independent posterior distributions, given by $p_i \sim \text{Beta}(X_i,N_i-X_i)$, $i = 1,2$.
Using the two beta posterior distributions,  the estimated VE and its credible intervals can be calculated by numerical methods with random sampling of the two beta distributions  (see attached computer program).  As a comparison to the results in  \cite{polack2020safety}, Table \ref{tab:model2res} presents the reported credible intervals using the alternative model. The first credible interval (90.8, 97.9) in the table is slightly shorter than the one (90.3, 97.6) in the paper, by a unit of 0.2. More importantly, this model presents the posterior distributions of $p_1$ and $p_2$, the two infection rates for the vaccine and placebo, as shown in Figure \ref{fig:ir-dis}.  

\begin{threeparttable}
	\centering
	\caption{Vaccine Efficacy with the independent priors of $p_1$ and $p_2$.} \label{tab:model2res}
	\begin{tabular}{p{16em}ccccc}
		\hline
		\textbf{Efficacy End Point} & \multicolumn{2}{p{8em}}{\textbf{BNT162b2}} & \multicolumn{2}{p{8em}}{\textbf{Placebo}} & \multicolumn{1}{p{10em}}{\textbf{Vaccine Efficacy, \% \newline{} (95\% Credible \newline{} Interval)}} \\
		\multicolumn{1}{c}{\textcolor[rgb]{ .141,  .125,  .129}{}} & \multicolumn{1}{p{4.02em}}{No. of\newline{}Cases\newline{}($X_1$)} & \multicolumn{1}{p{5.085em}}{No. of\newline{}Participants\newline{}($N_1$)} & \multicolumn{1}{p{4.02em}}{No. of\newline{}Cases\newline{}($X_2$)} & \multicolumn{1}{p{5.085em}}{No. of\newline{}Participants\newline{}($N_2$)} &  \\\hline
		\rowcolor{Gray}
		$\textVE_1:$ Covid-19 occurrence at least\newline{}
		\hspace*{2em}7 days after the second dose\newline{}
		\hspace*{2em}in participants without evidence \newline{}
		\hspace*{2em}of infection & 8     & 18198 & 162   & 18325 &  95.0\tnote{*}   (90.8, 97.9) \\\hline
		$\textVE_2:$ Covid-19 occurrence at least\newline{}
		\hspace*{2em}7 days after the second dose\newline{}
		\hspace*{2em}in participants with and those\newline{}
		\hspace*{2em}without evidence of infection & 9     & 19965 & 169   & 20172 &  94.6\tnote{*}   (90.4, 97.6) \\
		\hline
	\end{tabular}
	\begin{tablenotes}
		\item[*] Under the alternative model, the posterior means of VE are also 95.0 and 94.6  for the vaccine and placebo groups, respectively.  
	\end{tablenotes}
\end{threeparttable}


\end{document}